\documentclass[journal,subeqn,caption]{IEEEtran}

\usepackage{mathtools}
\usepackage{cite}
\usepackage{float}
\usepackage{multicol}
\usepackage{multirow}
\usepackage{hyperref}
\usepackage{graphicx}
\usepackage{amsfonts}
\usepackage{amsmath}
\usepackage{color}
\usepackage{epstopdf}
\usepackage{caption}
\usepackage{subcaption}
\usepackage{mathtools, nccmath}
\usepackage{multirow}
\usepackage{bbm}
\usepackage{marginnote}
\usepackage{tikz}
\usepackage{pbox}
\usetikzlibrary{arrows,decorations.pathmorphing,backgrounds,fit,positioning,shapes.symbols,chains, shapes,snakes}
\usepackage{xcolor}
\usepackage{comment}
\usepackage{amsthm}

\newtheorem{lemma}{Lemma}

\newcommand{\highlight}[1]{%
  \colorbox{gray!30}{$\displaystyle#1$}}

\newcommand{\specialcell}[2][c]{%
  \begin{tabular}[#1]{@{}c@{}}#2\end{tabular}}
\usepackage[font={small}]{caption}

\begin{document}

\title{Chance Constraints for Improving the Security of\\ AC Optimal Power Flow}

\author{M. Lubin, Y. Dvorkin, \textit{Member}, \textit{IEEE}, and L. Roald, \textit{Member}, \textit{IEEE}.
}

\maketitle

\begin{abstract}
This paper presents a scalable method for improving the solutions of AC Optimal Power Flow (AC OPF) with respect to deviations in predicted power injections from wind and other uncertain generation resources. 
The focus of the paper is on providing solutions that are more robust to short-term deviations, and that optimize both the initial operating point and a parametrized response policy for control during fluctuations. We formulate this as a chance-constrained optimization problem. To obtain a tractable representation of the chance constraints, we introduce a number of modelling assumptions and leverage recent theoretical results to reformulate the problem as a convex, second-order cone program, which is efficiently solvable even for large instances. Our experiments demonstrate that the proposed procedure improves the feasibility and cost performance of the OPF solution, while the additional computation time is on the same magnitude as a single deterministic AC OPF calculation.
\end{abstract}

\section{Introduction}%

Optimal power flow (OPF) is an integral part of many operational decision support tools in transmission system operation. Typical OPF formulations aim to optimize dispatch settings of controllable generators given their operating limits, forecasted operating conditions, power flow and nodal voltage constraints, and security margins. 
The recent push toward integrating  renewable energy resources with intermittent outputs has introduced a new degree of uncertainty and complexity in transmission operations. First, it requires dealing with nonconvex  and nonlinear AC power flow equations, which makes even the deterministic OPF problem  NP-hard \cite{pascalNPhard}. Second, it is difficult to model uncertainty propagation throughout the network.
One approach to circumvent those challenges is to replace the AC power flow equations with the 
linear DC approximation, which neglects power losses, assumes small angle differences, and parameterizes the voltage magnitudes. The linearity and convexity of the DC approximation enables the application of scenario-based \cite{867521}, chance-constrained \cite{ bienstock_2014} and robust  \cite{6917060,6948280,6575173} optimization techniques to deal with the uncertainty of renewable generation resources in a tractable manner.


Recent efforts to solve an uncertainty-aware AC OPF include \cite{6652374, 7314992, louca_2017, molzahn2018, roald2017, 7828060}. Vrakopoulou \textit{et al.} \cite{6652374} describe a chance constrained  OPF (CC-OPF) model based on a convex AC power flow relaxation and a scenario sampling approach to represent the uncertainty of nodal power injections. The model in 
\cite{7314992} is a two-stage robust AC OPF model that exploits convex relaxations of the inner problem. While the two above methods relax the power flow equations and hence cannot provide robust guarantees, \cite{louca_2017} devises an \emph{inner} approximation for the robust AC OPF, at the expense of assuming controllable injections at every node. In \cite{molzahn2018}, convex relaxations to provide conservative estimates on the impact of uncertainty for each constraint, thus guaranteeing robust feasibility of engineering limits for any uncertainty realization. The CC-OPF model in \cite{roald2017}  uses the full AC power flow equations for the forecasted operating point, while the impacts of uncertainty are modeled using linearized AC power flow equations. The CC-OPF for distribution networks in \cite{7828060} also models linearized  AC power flow equations, but does not enforce flow limits. 


While many of the above mentioned papers have been shown to provide good results on test cases, the techniques are frequently prohibitively expensive for large instances (e.g., because they use semidefinite programming), while still lacking rigorous guarantees on solution feasibility.
Since rigorous guarantees appear prohibitively conservative and hard obtain in practice, this paper focuses on computational tractability and proposes an approximate, scalable AC CC-OPF method.
Our formulation leverages a deterministic AC OPF solution, which is typically already available in practice, 
and then uses a chance-constrained formulation to robustify the solution against uncertain nodal power injections.  
To obtain a tractable AC CC-OPF formulation, we introduce a number of modelling assumptions. First, we suggest affine response policies to model the real-time response to power injection uncertainty. Second, we linearize the AC power flow equations around the deterministic AC OPF solution. 
Finally, we reformulate the chance constraints into convex second-order cone (SOC) constraints.  Relative to  previous studies, e.g.  \cite{7828060},  we also enforce chance constraints on apparent power flows, which quadratically depend on the uncertainty, using the recent result for quadratic chance constraints in \cite{LubinTwoSided}. The resulting AC CC-OPF is a SOC program (SOCP) with a similar number of constraints as the deterministic AC OPF; such SOCPs have shown to be tractable even for large instances \cite{bienstock_2014, roald2017}.
The numerical experiments demonstrate that the proposed AC CC-OPF is on par with  the deterministic AC OPF in terms of its computational performance, but has superior solution feasibility and cost performance. The ability to optimize the response policy also improves operational performance.

\section{CC AC-OPF Formulation} 
\label{sec:cc_ac_opf_full}
This section formulates an AC~CC-OPF starting from a standard deterministic AC OPF. 
To this end, we first review necessary preliminaries and then state the deterministic AC OPF model. To model the impact of uncertainty from renewable energy sources, we introduce a model for power injection uncertainty, as well as general response policies that would compensate for a power mismatch. 
We then state the AC~CC-OPF, and discuss the challenges related to solving the problem. 

\subsection{Preliminaries}

\subsubsection{Power injections} Let $\mathcal{N}$ denote the set of nodes with $m= |\mathcal{N}|$, and let vector $p_{G}$ indexed as $p_{G,i}$ for $i \in \mathcal{G}$ denote the total active power output of conventional generators at every node $i$. \textcolor{black}{For simplicity of notation, it is assumed that there is one generator per node, such that $\mathcal{G}=\mathcal{N}$. Nodes without generation or with more than one generator can be handled by setting the limits to zero or by changing notations, respectively; both modifications  do not  require changes in the proposed method. }
Similarly, $p_{D}$ indexed as $p_{D,i}$ denotes the total active power demand at every node, and the active power injections from renewable energy generators is denoted by the vector $p_{U}$ indexed as $p_{U,i}$. The corresponding reactive power injections are denoted by $q_{G}$, $q_{D}$, $q_{U}$. Thus, the vectors of net active and reactive power injections at all nodes are given by:
\begin{subequations}
\label{p_eq_det} 
\begin{align}
& p = p_G - p_D + p_U  \\  & q = q_G - q_D + q_U. 
\end{align}
\end{subequations}
In the following, we assume there is no curtailment of undispatchable renewable generation, i.e. $ p_U$ is a function solely of available wind and solar irradiation, and that all loads $p_D$ are fixed. These assumptions can be relaxed by modeling renewable curtailment, load shedding or demand response.   

\subsubsection{Nodal voltages} Vectors $v$ and $\theta$ indexed as $v_i$ and $\theta_i$ stand for the voltage magnitudes and voltage angles, respectively, at every node $i \in \mathcal{N}$. The range of acceptable voltage magnitudes is defined as   $v \in \big[v^{min},v^{max}\big]$. 

\subsubsection{Power flows} 
Let $\mathcal{L}$ denote the set of tuples $ij$, such that there is a line between node $i$ and node $j$. Note that we treat both $ij$ and $ji$ as distinct elements of $\mathcal{L}$, and for simplicity, we assume \textcolor{black}{that there is not more than one line connecting two nodes}. If there is more than one line connecting node $i$ and $j$ \textcolor{black}{(e.g. multi-circuit lines)}, they can be equivalently converted in one line. Note that $2l=|\mathcal{L}|$, where $l$ is the number of physical lines.
The vectors $f^p$ and $f^q$ indexed as $f^{p}_{ij}$ and $f^{q}_{ij}$ then denote the active and reactive power flows from node $i$ to node $j$ along line $ij$. Note that the power flows $f^{p}_{ij} \neq f^{p}_{ji}$ and $f^{q}_{ij} \neq f^{q}_{ji}$, due to line power losses. The apparent power flow limits are given by the vector  $s^{max}$ indexed as $s_{ij}^{max}$. 

The active and reactive power flows on each transmission line $ij \in \mathcal{L}$ from node $i$ to node $j$ depend nonlinearly on the voltage magnitudes $v$ and voltage angles $\theta$ at these nodes:
\begin{subequations}
\label{ac_pf_eq1}
\begin{align}
f^p_{ij}(v, \theta) =  v_i v_j \big[G_{ij} \cos (\theta_i - \theta_j) + B_{ij}  \sin (\theta_i - \theta_j)\big]    \label{ac_pf_eq1p}\\
f^q_{ij} (v, \theta) = v_i v_j \big[G_{ij} \sin(\theta_i - \theta_j) - B_{ij}  \cos (\theta_i - \theta_j)\big], \label{ac_pf_eq1q}
\end{align}
\end{subequations}
where $G_{ij}$ and $B_{ij}$ are the real and imaginary parts of the network admittance matrix $Y=G+jB$. The system is balanced when the power flows leaving each node is equal to the net injection at that node:

\begin{subequations}
\label{ac_pf_eq2}
\begin{align}
p_i = v_i^2G_{ii} + \sum_{j : ij \in \mathcal{L}} f^p_{ij}(v, \theta),  \label{eq:ac_pf_eq2_1}\\
q_i = -  v_i^2 B_{ii}  + \sum_{j : ij \in \mathcal{L}} f^q_{ij} (v, \theta),\label{eq:ac_pf_eq2_2}
\end{align}
\end{subequations}
where the first term represent the contributions from the nodal shunt elements. In the following, we will use $F\left( p, q, v, \theta\right) =0$ to denote the nodal power balance equations \eqref{p_eq_det} - \eqref{ac_pf_eq2}. 

\subsubsection{$pq,~pv$ and $\theta v$ nodes} As follows from \eqref{ac_pf_eq2}, there are four variables $p, q, v, \theta$ and two equations per node. This implies that only two of the four variables $p, q, v, \theta$ can be chosen independently, while the others are implicitly determined through \eqref{ac_pf_eq2}. In typical power system operations, there are three types of nodes, namely $pq,~pv$ and $\theta v$ nodes. The $pv$ and $pq$ nodes are characterized as nodes that maintain constant values of active power and voltage magnitude $(p_{i},v_i)$ and active and reactive power $(p_{i},q_{i})$, respectively. The $pv$ nodes are typically generation nodes, where the generators control their reactive power output to maintain constant voltage magnitudes $v_i$. The $pq$ nodes are any node where the power injections are directly specified, such as load nodes or nodes without generation or load. 
The $\theta v$ node is referred to as the reference or slack bus. At this node, the voltage angle and magnitude $(\theta_i,v_i)$ are kept constant, while the active and reactive power injections are implicitly determined. 
Note that any bus can have a combination of load, generation and renewable power. 

\subsubsection{Production cost} The production cost of each generator is defined as $c_i(p_{G,i})$ and is typically well approximated by a convex (e.g. linear or convex quadratic) function \cite{wood_2012}.




\subsection{Deterministic AC OPF}
\textcolor{black}{The standard deterministic AC OPF problem, as found in e.g. \cite{4075418},} can be formulated as follows:
\begin{subequations}
\label{detACOPF}
\begin{align}
\min_{\substack{p_G, q_G, \\v, \theta}} ~~&  \sum_{i\in \mathcal{N}} c_i(p_{G,i})&& \label{ac_obj}\\
\text{s.t.}  ~~         
&F\left( p, q, v, \theta \right) = 0, \label{ac_powerbal}  \\
&p_{G,i}^{min} \leq p_{G,i} \leq p_{G,i}^{max}, &&\forall {i\in\mathcal{N}}  \label{ac_p}\\
&q_{G,i}^{min} \leq q_{G,i} \leq q_{G,i}^{max}, &&\forall {i\in\mathcal{N}}  \label{ac_q}\\
&v_{j}^{min} \leq v_{j} \leq v_{j}^{max}, &&\forall {j\in\mathcal{N}}  \label{ac_v}\\
&(f^{p}_{ij}(\theta, v))^2\!\!+\!\!  (f^{q}_{ij}(v, \theta))^2 \!\!\leq\!\! (s_{ij}^{max})^2, &&\forall {ij\in\mathcal{L}} \label{ac_s}  \\
& \textcolor{black}{\theta_{ref} = 0}.  &&\label{ac_slack}
\end{align}
\end{subequations}
Eq.~\eqref{ac_obj} minimizes the total operating cost defined as the sum of production costs of all conventional generators. Eq.~\eqref{ac_powerbal} enforces the active and reactive nodal power balance as defined in \eqref{ac_pf_eq1}-\eqref{ac_pf_eq2}.
Eq.~\eqref{ac_p}-\eqref{ac_q} impose limits on the active and reactive power output of conventional generators. The nodal voltages and apparent power flow limits are limited in Eq.~\eqref{ac_v}-\eqref{ac_s}. Eq.~\eqref{ac_slack} sets the voltage angle at the reference node. 

The AC OPF in \eqref{detACOPF} returns the generation set points $p_{G}, q_G$ and voltage magnitudes $v$ that are to be fixed during steady-state operations at $pv$ nodes. As typical for $pq$ nodes, active and reactive power demands are  also fixed in \eqref{detACOPF}.

\subsection{Chance-Constrained AC Optimal Power Flow}
The AC OPF in \eqref{detACOPF} assumes that the power injections are perfectly known at the time of scheduling. In practice, this is not true, as both load and renewable generation might vary from their forecasted value. To ensure secure operation, it is therefore important to account for the impact of uncertainty on system operation. This paper will consider variations only in the renewable energy production. However, the method can be extended to incorporate other types of intermittent injections, such as load or distributed energy resources.

Assume that $\omega$ is a vector of real-time deviations of each renewable generator from its forecasted active power output $p_U$, such that the real-time renewable power production is given by ${p}_{U}(\omega) = p_U + \omega$.
The variation in active power is driven by variations in the primary energy source, such as the current wind speed or solar irradiation. The reactive power output corresponding to the same uncertainty realization $q_U(\omega)$ is typically linked to the active power generation through a set of grid requirements, e.g. maintaining a given constant power factor or contributions to local voltage control. 

As the power injections from renewable generators change, the controllable generators must adapt their active and reactive power $p_G(\omega)$ and  $q_G(\omega)$ to the uncertainty realization $\omega$. In particular, the response from the generators must ensure that the power injections $p_U(\omega), q_U(\omega), p_G(\omega), q_G(\omega)$ yield a feasible power flow solution, i.e., one that satisfies  power flows in \eqref{ac_powerbal} and determines nodal voltages  $v(\omega)$ and $\theta(\omega)$. In addition, the generation response $p_G(\omega),~q_G(\omega)$ must be chosen to respect the power output limits \eqref{ac_p}-\eqref{ac_q}, and satisfy constraints on the voltage magnitudes \eqref{ac_v} and power flows \eqref{ac_s}.

Without assuming a specific generation response, the AC CC-OPF can be formulated as:

\begin{subequations}
\label{originalACCCOPF}
\begin{align}
&\!\!\!\!\!\!\!\! \min_{\substack{q_U(\cdot), p_G(\cdot), \\ q_G(\cdot), v(\cdot), \theta(\cdot)} } ~~ \sum_{i\in \mathcal{N}} \mathbb{E} \left[c_i (p_{G,i} (\omega))\right] && \label{CCac_obj}\\
&\text{s.t.}  ~ \forall_{i\in\mathcal{G}}, ~\forall_{j\in\mathcal{N}}, ~\forall_{ij\in\mathcal{L}}    \nonumber     \\
&\quad F\left( {\theta}(\omega),  {v}(\omega),  {p}(\omega), {q}(\omega) \right) = 0 ~\forall_{\omega} \label{CCac_powerbal}\\
&\quad \mathbb{P}( {p}_{G,i}(\omega) \leq p_{G,i}^{max})\geq 1-\epsilon_P, \label{CCac_pMax}\\
&\quad \mathbb{P}( {p}_{G,i}(\omega) \geq p_{G,i}^{min})\geq 1-\epsilon_P,  \label{CCac_pMin}\\
&\quad \mathbb{P}( {q}_{G,i}(\omega) \leq q_{G,i}^{max})\geq 1-\epsilon_Q,  \label{CCac_qMax}\\
&\quad \mathbb{P}( {q}_{G,i}(\omega) \geq q_{G,i}^{min})\geq 1-\epsilon_Q,  \label{CCac_qMin}\\
&\quad \mathbb{P}( {v}_{j}(\omega) \leq v_{j}^{max})\geq 1-\epsilon_V, \label{CCac_vMax}\\
&\quad \mathbb{P}( {v}_{j}(\omega) \geq v_{j}^{min})\geq 1-\epsilon_V, \label{CCac_vMin}\\
&\quad \mathbb{P}( (f^{p}_{ij}(\omega))^2+ (f^{q}_{ij}(\omega))^2  \leq (s_{ij}^{max})^2)\geq 1-\epsilon_I, \label{CCac_iMax}\\
&\quad \theta_{ref}(\omega) = 0. \label{CCac_ref} 
\end{align}
Here, expectations $\mathbb{E}$ and probabilities $\mathbb{P}$ are defined over the distribution of $\omega$. The nodal power injections $p(\omega)$ and $q(\omega)$ are derived from \eqref{p_eq_det} as:
\label{p_eq_cc}
\begin{align}
& p (\omega) = p_G (\omega)  - p_D  + p_U  (\omega)  \\  & q (\omega)  = q_G (\omega)  - q_D   + q_U  (\omega), 
\end{align}
\end{subequations}
and the active and reactive power flows $f^p(\omega) = f^p(v(\omega),\theta(\omega))$ and $f^q(\omega) = f^q(v(\omega),\theta(\omega))$ are as in \eqref{ac_pf_eq1}. 

Eq.~\eqref{CCac_obj} minimizes the expected total operating cost, including the cost of serving the forecasted demand and responding to deviations $\omega$. Eq.~\eqref{CCac_powerbal} ensures that generation response is chosen such that the nodal power balance holds for any realization of uncertainty and system response, i.e. the solution does not tolerate or assume availability of load shedding. The limits on the power output of conventional generators, voltage magnitudes and apparent power flows are enforced using the separate chance constraints \eqref{CCac_pMax}-\eqref{CCac_iMax}. The chance constraints require that the constraint should hold with a prescribed (typically high) probability.
The risk level associated with the chance constraint can be regulated by the choice of the violation probabilities
$\epsilon_P, \epsilon_Q, \epsilon_V, \epsilon_I$.  
As in the deterministic model, the voltage angle is set to zero at the reference node \eqref{CCac_ref}.

The AC CC-OPF as formulated in \eqref{originalACCCOPF} cannot be solved using known solution strategies. First, it inherits the nonlinear and nonconvex properties from the AC power flow equations, which must be shown to have a solution for all $\omega$. Second, it relies on generic response policies which gives rise to an infinite set of both decision variables and constraints, and is hence an infinite-dimensional optimization problem. In addition, even in the case where the uncertain power injections $\omega$ follow a Gaussian distribution, it is challenging to derive the statistics of the output variables (power flows, voltages and generation outputs) due to the nonlinearity of the power flow equations. Therefore, the chance constraints~\eqref{CCac_pMax}-\eqref{CCac_iMax} are not known to be tractable in general. Section~\ref{sec:cc_our_method} presents a number of simplifying, yet practically feasible, assumptions  to overcome our inability to solve  \eqref{originalACCCOPF}. 

Other AC CC-OPF approaches, e.g. \cite{6652374, BAKER2017230}, exploit joint chance constraints, which limits the probability that any of the constraints are violated, instead of the separate chance constraints, which limit the violation probability only for individual constraints. Our choice of separate chance constraints is motivated as follows. First, separate chance constraints can be viewed as more appropriate for power system operations, as they limit the risk of individual component failures, thus pointing to particular high-risk components or areas. Second, while the joint constraints provide only weak guarantees on violation probability (e.g., via the Bonferroni approximation~\cite{XieThesis})
previous OPF studies \cite{roald2017} have shown that separate chance constraints limit the joint violation probability effectively due to the low number of active constraints. Finally, the joint chance constraints are notoriously difficult to enforce. Existing approaches with joint feasibility guarantees are either overly conservative, e.g. in \cite{6652374, BAKER2017230, roald2017}, or computationally demanding \cite{XieThesis}.

\section{A Tractable AC CC-OPF Formulation} \label{sec:cc_our_method}

This section presents modeling choices  to obtain a computationally tractable approximation of the AC CC-OPF \eqref{originalACCCOPF}. 

\subsection{System Response}

To overcome the infinite dimensionality of \eqref{originalACCCOPF}, we develop a family of finitely parameterized response policies $q_U(\omega), p_G(\omega), q_G(\omega), v(\omega), \theta(\omega)$. As described below, some of these policies can be defined  explicitly with respect to $\omega$, while some can only be formalized using  the implicit constraints dictated by the AC power flow model~\eqref{ac_powerbal}.  

\subsubsection{Renewable Reactive Power Generation}
The active and reactive power outputs of wind power generators are inherently related. While different grid operators have different requirements on the reactive power control from renewable generators, we adopt a common approach to maintain a constant power factor $\cos(\phi)$, in which the reactive power output will change following the deviation of the active power output:
\begin{eqnarray}
	{q}_{U,i}(\omega)= \gamma_i p_{U,i} (w) = \gamma_i p_{U,i}  + \gamma_i \omega_i \label{gen_response2},
\end{eqnarray}
where $\gamma_i = \sqrt{\left(1-\cos^2 \phi_i\right)}/\cos \phi_i$. Although the valid physical range $\cos \phi_i$ is $[-1,1]$, it is typical that $\cos \phi_i \approx 1$.  {\color{black} The value of $\gamma_i$ in \eqref{gen_response2} can be either  optimized, if the operator is able to control the power factor in real-time, or fixed ahead of time, if otherwise. In the following, $\gamma_i$ is optimized since it is a more general case, which can be adjusted  to model fixed $\gamma_i$ as a special case. 

More general, relative to  \eqref{gen_response2},   relationships between the active and reactive power fluctuations can  be considered by introducing separate random variables to represent  reactive power fluctuations.}  

\subsubsection{Generation and Voltage Control}
Following fluctuations $\omega$, the controllable generators adjust their reactive and active power outputs to ensure power balance and maintain the desired voltage levels. {\color{black}The balancing policy described here is similar to standard approaches in power system operations and have been adapted (with some modifications) from \cite{vrakopoulou2013TPWRS, roald2017}.

For the purposes of this paper, we assume that active power is balanced by activation of reserves, imitating the Automatic Generation Control (AGC).  
The total power mismatch \textcolor{black}{$\Omega=\sum_{i\in\mathcal{N}} \omega_i$} due to forecast errors is split among generators based on  participation factors $\alpha$ based on the following generation control policy:
\begin{align}
 	p_{G,i}(\omega) = p_{G,i} - \alpha_i \Omega. \label{policy_reserve_CC}
\end{align}
In this paper, we optimize the participation factors $\alpha$ along with the scheduled power generation. However, a simpler case with fixed $\alpha$ could also be considered. 
To ensure that a given mismatch is balanced by the same amount of reserve activation, the participation factors are required to sum to 1:
\begin{subequations}
 \label{eq:reserveCC}
\begin{align}
 	\sum_{i\in\mathcal{G}} \alpha_i = 1.  && \label{eq:reserveCC3}
\end{align}
Since we assume that the generation control policy \eqref{policy_reserve_CC} represents the activation of reserves, it is natural to introduce a new set of optimization variables $r_i$ which represent the reserve capacity from each generation $i\in\mathcal{G}$. For simplicity, we consider the reserve capacity assignment to be symmetric, i.e. the same capacity $r_i$ is scheduled for both up- and down-regulation, although this could easily be generalized. To ensure sufficient reserves to cover  $-\alpha\Omega$ with a high probability, we enforce the following chance constraints:
\begin{align}
   \hspace{-3mm}   & \mathbb{P}[-\alpha_i\Omega \leq r_i] \geq 1\!-\! \epsilon_P, ~
    &&   \hspace{-3mm}  \mathbb{P}[-\alpha_i\Omega \geq -r_i] \geq 1\!-\! \epsilon_P, \forall i\in\mathcal{G} \label{eq:reserveCC2}
\end{align}
where the left-hand side represents the power mismatch and the right-hand side describes the available reserve capacity. 

Finally, we must ensure that the scheduled generation set-points $p_{G,i}$ are such that the reserves $r_i$ can be delivered without violating the upper and lower generation bounds. This is done by enforcing 
\begin{align}
  \hspace{-3mm}  & p_{G} + r \leq p_{G}^{max}, ~     &&   \hspace{-3mm}  p_{G} - r \geq p_{G}^{min}.  \label{eq:reserveCC1}
\end{align}
\end{subequations}
}
The policy in \eqref{policy_reserve_CC} only balances the power mismatch due to forecast errors $\omega$. Due to power flows and voltages  changes following any realization $\omega$, the active power losses in the system will also change. This \emph{change} in the power losses is 
typically small relative to the losses at the forecasted operating point, and we assume that any changes to the power losses will be balanced via  the $\theta v$ bus. 

For reactive power balancing and voltage control, a distinction between $pv,~pq$ and $\theta v$ buses becomes significant. Considering common practice, we assume that the reactive power injections are constant at $pq$ buses, while generators at $pv$ and $\theta v$ buses adjust their reactive power outputs to keep the voltage magnitude constant. Note that a centralized voltage control scheme as in \cite{6652374}, where the reactive power mismatch is distributed among generators according to optimized participation factors, could also be implemented.

\subsubsection{Explicit and Implicit Nodal Response Policies}
Some optimization variables in the AC CC-OPF are explicitly related  to the initial dispatch $p_G, q_G$, the voltage magnitudes $v$ and the response policy parameters $\gamma$ and $\alpha$. In particular, the active and reactive power injections at $pv$ buses, the active power injection and voltage magnitude at $pv$ buses and the voltage angle and magnitude at $\theta v$ buses remain constant for any  $\omega$. On the other hand, some variables are not directly controlled, but rather implicitly determined through the AC power flow equations. This holds for the active and reactive power $p_{G}(\omega),~q_G(\omega)$ at the $\theta v$ bus, the reactive power $q_G(\omega)$ and voltage angle $\theta(\omega)$ at the $pv$ buses and the voltage magnitude and angle $v(\omega),~\theta(\omega)$ at $pq$ buses. 

Table~\ref{table:response_pol} summarizes the variables that are explicitly defined by the response policies for each node type, and shows the variables that are only implicitly determined. Relative to the generic response policies in \eqref{originalACCCOPF}, the response policies in Table~\ref{table:response_pol} depend on a finite number of decision variables, namely the initial dispatch $p_G, q_G$, the voltage magnitudes $v$ and the response policy parameters $\gamma$ and $\alpha$. Introducing these response policies yields an optimization problem with a finite number of decision variables. 

\begin{table}
\begin{center}
        \captionsetup{justification=centering, labelsep=period, font=small, textfont=sc}
\caption{Explicit and Implicit (impl.) Response Policies}
\normalsize
\label{table:response_pol}
\begin{tabular}{ c| c c c }
\hline
\hline
\multirow{2}{*}{\specialcell{Response \\ Policy}}& \multicolumn{3}{c}{Node type}\\
\cline{2-4}
 &$pv$ & $pq$ & $\theta v$ \\ 
 \hline
 \hline
$p_{U,i} (\omega)$      &$p_{U,i} + \omega_i $                                   & $p_{U,i} + \omega_i $ & $p_{U,i} + \omega_i $  \\
$q_{U,i} (\omega)$      & $q_{U,i} + \highlight{\gamma_i} \omega_i $  & $q_{U,i} + \highlight{\gamma_i} \omega_i $  & $q_{U,i} + \highlight{\gamma_i} \omega_i $ \\  
$p_{G,i} (\omega)$      & $ \highlight{p_{G,i}}-\highlight{\alpha_i} \Omega $    & $\highlight{p_{G,i}}-\highlight{\alpha_i} \Omega $ & impl.  \\  
$q_{G,i} (\omega)$      & impl. & $\highlight{q_{G,i}}$                                     & impl. \\ 
$\theta_{i} (\omega)$   & impl. & impl.                                             & 0   \\
$v_{i} (\omega)$        & $\highlight{v_i}$                                     & impl.  & $\highlight{v_i}$ \\
\hline 
\hline
\end{tabular}
\end{center}
Grey background denotes the optimization variables of the AC CC-OPF. 

\end{table}

\subsection{Linearization of AC Power Flow Equations} \label{sec:ac_pf_eq_linearization}
The implicit values in Table~\ref{table:response_pol} are determined through the AC power flow equations, which are nonlinear and do not permit an explicit solution. Even guaranteeing that a solution exists for a range of power injections is an open research topic \cite{molzahn2018}. 
Therefore, this paper aims to obtain a more robust solution than the deterministic AC OPF, while maintaining computational efficiency and scalability, rather than to provide a comprehensive theoretical guarantees for the probabilistic constraints and for AC power flow solvability. This motivates to linearize the nodal power balance equations $F(p,q,v,\theta) = 0$ around the forecasted operating point. This linear model, in combination with the generation control policy described above, enables us to replace \eqref{CCac_powerbal} by a set of linear constraints and to obtain explicit analytical expressions for the implicit response policy.

Since $F$ is a smooth function, we can define its first order Taylor expansion, or linearization, at a point $(\bar p, \bar q, \bar v, \bar \theta)$ as
\begin{gather}
\bar F(p,q,v,\theta; \bar p, \bar q, \bar v, \bar \theta) = \nonumber \\F(\bar p, \bar q, \bar v, \bar \theta) + J_F(\bar p, \bar q, \bar v, \bar \theta)((p,q,v,\theta)-(\bar p, \bar q, \bar v, \bar \theta)) \label{ac_lin_jacob},
\end{gather}
where $J_F$ is the Jacobian matrix of $F$ at the given point. \textcolor{black}{ We also define the line flows as analogous linearized functions $\bar f^p(v,\theta; \bar v, \bar \theta)$ and $\bar f^q(v,\theta; \bar v, \bar \theta)$. The use of the linearization \eqref{ac_lin_jacob} is motivated by its connections with the Karush-Kuhn-Tucker (KKT) optimality conditions elaborated in the following lemma.} 

 \begin{lemma}\label{kktlemma}
 Let $(\bar p_G, \bar q_G, \bar v, \bar \theta)$ be a locally optimal solution to the non-linear, deterministic AC OPF~\eqref{detACOPF}, i.e., it satisfies the first-order optimality conditions of Theorem 12.1 of~\cite{NoceWrig06}. Then $(\bar p_G, \bar q_G, \bar v, \bar \theta)$ is a globally optimal solution to the following problem with the linearized AC power flow constraints:
 \begin{subequations}

\label{detlinACOPF}
\begin{align}
\min_{\substack{p_G, q_G, v, \\\theta}} ~&  \sum_{i\in \mathcal{N}} c_i(p_{G,i})&& \label{ac_linobj}\\
\text{s.t.}  ~~         
&\bar F\left( p, q, v, \theta; \bar p, \bar q, \bar v, \bar \theta \right) = 0, \label{ac_linpowerbal}  \\
&(\bar f^{p}_{ij}(v, \theta; \bar v, \bar \theta))^2+ (\bar f^{q}_{ij}(v, \theta; \bar v, \bar \theta))^2 \leq  \nonumber \\ & ~~~~~~~~~~~~~~~~~~~~~~(s_{ij}^{max})^2, \forall {ij\in\mathcal{L}} \label{ac_lins} \\
&\eqref{ac_p},\eqref{ac_q},\eqref{ac_v},\eqref{ac_slack}.
\end{align}

\end{subequations}

\begin{proof}
The KKT conditions for \eqref{detACOPF} and \eqref{detlinACOPF} are identical, hence $(\bar p_G, \bar q_G, \bar v, \bar \theta)$ is a locally optimal solution of \eqref{detlinACOPF}. Furthermore, \eqref{detlinACOPF} is convex, so local optimality implies global optimality.
\end{proof}
 \end{lemma}
 
Hence, this linearization based on the first-order Taylor expansion does not perturb the optimality of $(\bar p_G, \bar q_G, \bar v, \bar \theta)$ for the forecasted system state where $\omega=0$. \textcolor{black}{This property is unique to the first-order Taylor expansion, to our knowledge. In the chance constrained problem~\eqref{approximateACCCOPF} later formulated,} the same objective is minimized over a subset of the feasible region of~\eqref{detlinACOPF}. \textcolor{black}{Hence, if this subset is not too restrictive,} the optimal solution is expected to remain relatively close to $(\bar p_G, \bar q_G, \bar v, \bar \theta)$. This is \textcolor{black}{an} intuitive but not fully rigorous justification for choosing this first-order Taylor expansion.

\textcolor{black}{Given this linearization,}
we then require that our response policy satisfies
\begin{equation}\label{eq:linearizedresponse}
    \bar F(p(\omega),q(\omega),v(\omega),\theta(\omega); \bar p, \bar q, \bar v, \bar \theta) = 0.
\end{equation}
Note that~\eqref{eq:linearizedresponse} consists of two equations for each node (from~\eqref{eq:ac_pf_eq2_1}-\eqref{eq:ac_pf_eq2_2}), while Table~\ref{table:response_pol} provides expressions for two implicitly defined values per node for any node type. Hence, the linear system~\eqref{eq:linearizedresponse} is well posed and yields a unique solution assuming $J_F(\bar p, \bar q, \bar v, \bar \theta)$ is invertible. As noted in \cite{bienstock_book},  $J_F (\cdot)$ is normally invertible for steady-state power grid conditions, with the exception of bifurcation points. One may use algorithmic differentiation~\cite{Griewank2008} to efficiently compute \textcolor{black}{$J_F$} and basic linear algebra to obtain the implicit response policies as explicit, affine functions of the explicit response policies analogously to~\cite{bienstock_2014}.

\textcolor{black}{Instead of using the first-order Taylor expansion to linearize the AC power flows in \eqref{ac_lin_jacob}, our method can use other linearization techniques, see \cite{coffrin_2014} for a review of such techniques. We experimented with using the fast decoupled load flow linearization~\cite{FastDecoupled}; this performed less well than the first-order Taylor expansion for recovering solutions that satisfy AC feasibility. }

\subsection{Chance Constraint Reformulation}

With the linearization of the AC power flow equations, we can express the generation outputs $p_{G,i}(\omega), q_{G,i}(\omega)$, the voltage magnitudes $v_i(\omega)$ and the active and reactive power flows $f^{p}_{ij}(\omega),~f^{q}_{ij}(\omega)$ as linear functions of the random deviations $\omega$. For the chance constraints on $p_{G,i}(\omega),~q_{G,i}(\omega)$ and $v_i(\omega)$ given by \eqref{CCac_pMax} - \eqref{CCac_vMin}, this linearity enables the use of well-known analytic chance constraint reformulations previously applied to the DC approximation \cite{bienstock_2014} and other AC linearizations \cite{roald2017, 7828060}. However, the power flow constraints \eqref{CCac_iMax} have a quadratic dependence on $\omega$, which has not been treated before. In the following, we first present the reformulation for the standard linear constraints and then extend the discussion to the quadratic chance constraints based on new results from \cite{LubinTwoSided}. For the derivation, we will assume that $\omega$ follows a Gaussian distribution with mean $\mu_\omega = 0$ and known covariance matrix $\Sigma_\omega$. However, these results are extendable to other known or partially known distributions (using distributionally robust optimization) as discussed below. 

\subsubsection{Chance constraints with linear dependence on $\omega$}
Under the assumption of Gaussianity, the chance constraints on \eqref{CCac_pMax}-\eqref{CCac_vMin},
\eqref{eq:reserveCC2} with a linear dependence on $\omega$ have an exact reformulation given by: 

\begin{subequations}
\label{chancesocexample}
\begin{align}
    &p_{G,i}^{min} \le \mathbb{E}[p_{G,i}(\omega)] \pm \Phi^{-1}(1-\epsilon_P)\operatorname{Stdev}[p_{G,i}(\omega)] \le p_{G,i}^{max}, \label{CCgen} \\
    &q_{G,i}^{min} \le \mathbb{E}[q_{G,i}(\omega)] \pm \Phi^{-1}(1-\epsilon_Q)\operatorname{Stdev}[q_{G,i}(\omega)] \le q_{G,i}^{max}, \label{CCq}\\
    &v_i^{min} \!\le\! \mathbb{E}[v_{i}(\omega)] \pm \Phi^{-1}(1-\epsilon_V)\operatorname{Stdev}[v_{i}(\omega)] \!\le\! v_i^{max},\! \label{CCv}
\end{align}
\end{subequations}
where $\Phi^{-1}$ is the inverse Gaussian cumulative distribution.
Because of the linear dependency on $\omega$, analytical expressions for the expectations and standard deviations are easy to obtain. 
Using the voltage magnitude $v_i$ as an example, the expectations are  $\mathbb{E}[v_{i}(\omega)]=v_{i}(\mu_\omega)=v_{i}(0)$, i.e.  linear in the decision variables, while the standard deviations are defined by: 
\begin{equation}
    \operatorname{Stdev}[v_{i}(\omega)] = \sqrt{J_{v_{i},\omega}(\alpha,\gamma)^\top \Sigma_\omega J_{v_{i},\omega}(\alpha,\gamma)},
\end{equation}
where $J_{v_{i},\omega}(\alpha,\gamma)$ is a vector of sensitivity factors describing the change in voltage magnitude $v_i$ as a function of the fluctuation $\omega$, derived from the AC power flow linearization and assumed generation control policies. These sensitivity factors are linear functions of variables $\alpha$ and $\gamma$. The reformulated chance constraints \eqref{chancesocexample} therefore representable as SOC constraints.

\subsubsection{Chance constraints with quadratic dependence on $\omega$} \label{sec:handling_cc_25_eps}
For the quadratic constraint~\eqref{CCac_iMax} no directly tractable reformulation is known. It is known to be convex only when $\epsilon_I$ is very small~\cite{vanAckooij2018}. Hence, we replace it with the inner approximation~\cite[Lemma 17]{LubinTwoSided}:
\begin{subequations}\label{quadapproxsplit}
\begin{align}
 &   \mathbb{P}(|f^{p}_{ij}(\omega)| \le t^p_{ij}) \ge 1 - \frac{\epsilon_I}{2}, \forall ij \in \mathcal{L}\label{abs1}\\
 &   \mathbb{P}(|f^{q}_{ij}(\omega)| \le t^q_{ij}) \ge 1 - \frac{\epsilon_I}{2}, \forall ij \in \mathcal{L}\label{abs2}\\
 &   (t^p_{ij})^2 +  (t^q_{ij})^2 \le (s_{ij}^{max})^2, \forall ij \in \mathcal{L}, \label{quadapproxsoc}
\end{align}
\end{subequations}
where $t^p_{ij}$ and $t^q_{ij}$ are auxiliary decision variables and~\eqref{quadapproxsoc} is a deterministic convex quadratic constraint.

We treat absolute value chance constraints in~\eqref{abs1} and~\eqref{abs2} using the SOC approximation developed in~\cite[Lemma 16]{LubinTwoSided}. \textcolor{black}{A direct application of this lemma (given that $\mathbb{E}[f_{ij}^p(\omega)] = f_{ij}^p(0)$ and $\mathbb{E}[f_{ij}^q(\omega)] = f_{ij}^q(0)$) implies that} \eqref{abs1} and~\eqref{abs2} may be inner approximated as:
\begin{subequations}\label{absapprox}
\begin{align}
    -t_{ij}^* - f_{ij}^*(0) &\le \Phi^{-1}(\frac{\epsilon_I}{2.5})\operatorname{Stdev}[f_{ij}^*(\omega)], \forall ij \in \mathcal{L} \\
    t_{ij}^* - f_{ij}^*(0) &\ge \Phi^{-1}(1-\frac{\epsilon_I}{2.5})\operatorname{Stdev}[f_{ij}^*(\omega)], \forall ij \in \mathcal{L}\\
    t_{ij}^* &\ge \Phi^{-1}(1-\frac{\epsilon_I}{5})\operatorname{Stdev}[f_{ij}^*(\omega)], \forall ij \in \mathcal{L}
\end{align}
\end{subequations}
for $*=p$ and $*=q$ respectively. The constraints~\eqref{quadapproxsplit} and~\eqref{absapprox} for $p$ and $q$ together imply that~\eqref{CCac_iMax} holds with probability $\epsilon_I$. \textcolor{black}{The inner approximation argument for constraints~\eqref{quadapproxsplit} is based on the union bound and hence results in the coefficients of 2 that appear in the denominators. The inner approximations of~\eqref{abs1}-\eqref{abs2} themselves introduce another factor of 1.25. $\epsilon_I$ is therefore multiplied by the inverse of 2.5 to obtain a provably conservative approximation. Omitting this factor of 2.5 would result in an \textit{outer approximation} of~\eqref{CCac_iMax}. In practice some factor between 1.0 and 2.5 could be chosen based on empirical tuning to balance the goal of satisfying the constraint with the target probability and the possible over-conservatism of the inner approximation used, in order to preserve feasibility of the AC CC-OPF. We did not investigate this possibility in this work.}

\subsubsection{Generalization beyond Gaussian distribution}
The assumption that $\omega$ follows a Gaussian distribution with known parameters can be relaxed by considering distributional robustness (i.e. partial knowledge about the distribution) in 
two  ways. Ref. \cite{rccopf, 7268773} and \cite[Lemma 8]{LubinTwoSided} discuss robustness with respect to the Gaussian parameters, while \cite{roald2015arxiv},~\cite{summers2015},~\cite{bowen2016},~\cite{8294298},~\cite{XieThesis}
discuss ambiguity in the type of distribution given 
known $\mu_{\omega}$ and $\Sigma_{\omega}$.
In both settings, a tractable SOCP formulation of the linearized AC CC-OPF can be obtained, hence enabling 
more general distributions without  compromising computational tractability.

\subsection{Joint Chance Constraints for the Reserve Capacities}
In \eqref{eq:reserveCC2}, we replaced the chance constraints on the  active power generation \eqref{CCac_pMax}, \eqref{CCac_pMin} with chance constraints on reserve capacity $r_i$, which only depend on the total power mismatch $\Omega$, i.e., a scalar random variable. We thus recast \eqref{eq:reserveCC2} as:
\begin{align}
    &\pm\Phi^{-1}(1-\epsilon_P)\operatorname{Stdev}[\alpha_i\Omega] = \nonumber \\
    &\pm\alpha_i \Phi^{-1}(1-\epsilon_P) \operatorname{Stdev}[\Omega] \le r_i, \quad \forall i\in\mathcal{G}
    \label{eq:CCreservesRef}
\end{align}
which is now a linear constraint in the decision variables $\alpha_i$ and $r_i$, since the standard deviation of the total fluctuation $\operatorname{Stdev}[\Omega]=\sqrt{\bold{1}^\top \Sigma_\omega \bold{1}}$ is a constant.
Furthermore, note that $\Omega_{1-\epsilon_P}=\Phi^{-1}(1-\epsilon_P) \operatorname{Stdev}[\Omega]$ is the $1-\epsilon_P$ quantile of $\Omega$. 
Hence, by enforcing \eqref{eq:CCreservesRef}, it follows that all the reserve chance constraints \eqref{eq:reserveCC2} will hold \emph{jointly} as long as $-\Omega_{1-\epsilon_P} \le \Omega \le \Omega_{1-\epsilon_P}$. The safety level $1-\epsilon_P$ in the reserve constraints hence have a natural interpretation as the probability of having sufficient reserve capacity available in the system. Furthermore, if we sum the reserve constraints for all generators, we obtain a total reserve capacity requirement:
\begin{align}
    \sum_{i\in\mathcal{G}} \alpha_i \Omega_{1-\epsilon_P} = \Omega_{1-\epsilon_P} \sum_{i\in\mathcal{G}} \alpha_i = \Omega_{1-\epsilon_P} \leq \sum_{i\in\mathcal{G}} r_i. \label{DETReserve}
\end{align}
This requirement is similar to probabilistic reserve requirements applied in, e.g., Switzerland \cite{7081775}, and can also be enforced within a deterministic AC OPF.


\subsection{Cost Function Approximation}
For simplicity, we replace the objective function \eqref{CCac_obj} with the deterministic value $\sum_{i\in \mathcal{N}} c_i (p_{G,i} (0))$. This is an exact reformulation when each $c_i$ is linear, given that $p_{G,i}(\omega)$ is an affine function of $\omega$, and we assume that $\omega$ follows a symmetric distribution with $\mathbb{E}[\omega] = 0$. In the more common case where each $c_i$ is a convex quadratic function,~\cite{bienstock_2014} show an exact reformulation for the Gaussian case. As this work focuses on feasibility more so than operational costs, we choose to use the deterministic cost approximation.

\textcolor{black}{Note that the objective function could also be extended to explicitly account for the cost of reserve provision, i.e., the remuneration of generators for maintaining reserve capacities $r_i$. Such models would only require a minor change in the objective function, for example as modelled in \cite{vrakopoulou2013TPWRS}.}

\subsection{A Tractable Approximation of AC CC-OPF}
\label{sec:tractableapprox}

Given the simplifications made in the previous sections, we can now state a tractable problem that approximates~\eqref{originalACCCOPF}.
We fix a linearization point $(\bar p, \bar q, \bar v, \bar \theta)$, and for all that follows, the implicit response policies are chosen uniquely to satisfy~\eqref{eq:linearizedresponse}. With this, our problem is given by:
\begin{subequations}
\label{approximateACCCOPF}
\begin{align}
& \min_{\substack{p_{G}, q_{G},  v \\ \alpha, \gamma, \theta }}  && \sum_{i\in \mathcal{N}} c_i (p_{G,i} (0))                            && \text{Expected Cost}\\
& \text{s.t}   && \eqref{eq:linearizedresponse} && \text{Linearized Power Flow} \\
&   && \eqref{eq:reserveCC1}, \eqref{eq:reserveCC3}, \eqref{eq:CCreservesRef}          & &\text{Active Power, Reserves} \\
&       && \eqref{CCq}, \eqref{CCv} \label{approxplaincc}                       && \text{Reactive Power, Voltage} \\
&       && \eqref{quadapproxsoc}, \eqref{absapprox}                             && \text{Apparent Power Flow}.
\end{align}
\end{subequations}
When the chance constraints are represented as SOC constraints 
and the objective is a convex quadratic function then \textcolor{black}{the approximate problem in \eqref{approximateACCCOPF} is solvable with standard SOCP methods that are used for many engineering applications and generally scale well \cite{LOBO1998193}.} The decision variables are those highlighted in Table~\ref{table:response_pol}, the reserve capacities $r_i$ and the auxiliary variables used to model the chance constraints.

The choice of the linearization point is an essential part of the formulation of~\eqref{approximateACCCOPF} even if its dependence is not explicit. Note that when the covariance of $\omega$ is exactly zero, the $\alpha$ and $\gamma$ variables become irrelevant, and the problem reduces to~\eqref{detlinACOPF}. Following Lemma~\ref{kktlemma}, the optimal solution to~\eqref{approximateACCCOPF} exactly matches that of the deterministic AC OPF, which is a desirable property that does not generally hold for other linearizations. 

\section{Solution Procedure}
Since the solution quality of \eqref{approximateACCCOPF} depends on the linearization point, we propose a stepwise solution procedure:
\begin{enumerate}
    \item Solve the deterministic AC OPF in \eqref{detACOPF}, with the reserve constraints \eqref{eq:reserveCC1} and capacity requirement \eqref{DETReserve}.
    \item Take this solution as the linearization point.
    \item Solve the  approximate AC CC-OPF in \eqref{approximateACCCOPF} to optimize generator set points and response policy.
\end{enumerate}
This solution approach has several advantages. First, it uses a standard deterministic AC OPF solver to obtain a feasible (albeit not robust to uncertainty) OPF solution. Second, if the uncertainty is relatively small, the linearization will provide good approximations to the full AC power flow equations. Third, it allows for analytical reformulation into
a problem with linear and convex SOC constraints, which can be solved using specialized solvers (e.g., Cplex, Gurobi, Mosek), dynamic linear outer approximations methods (e.g., \cite{bienstock_2014, rccopf}) or lazy constraint generation (e.g., \cite{roald2017corrective}). \textcolor{black}{Finally, if the AC CC-OPF at the solution procedure is infeasible, the acceptable violation probabilities. $\varepsilon_I, \varepsilon_P, \varepsilon_Q, \varepsilon_V$ may be increased to obtain a feasible, though less secure solution.}

In general, the approach can be expected to yield fast and high-quality solutions in cases where the optimal solution to the original nonlinear AC CC-OPF is not too far away from the deterministic AC OPF solution. 
\textcolor{black}{Similarly,  one can extend the formulation to account for contingencies, e.g., based on  \cite{roald2017corrective}.} 

\section{Case Study}

We use the 118-node IEEE test system \cite{matpower} with the following modifications. 
Parameters $s_{ij}^{max}$ are reduced by 20\%. The values of $p_{D,i}$ and $q_{D,i}$  are increased by 20\% at every node and parameters $v_{i}^{min}$ and $v_{i}^{max}$ at $pq$ nodes are set to 0.95 and 1.05 p.u., respectively. Parameters $q_{G,i}^{min}$ and $q_{G,i}^{max}$  are set to 90\% of their rated values. We include 11 wind farms with the total forecast power output of 1196 MW, itemized as in Table~\ref{table:data}, which is $\approx 28.2\%$ of the total active power demand. As in \cite{rccopf}, the wind power forecast error is  zero-mean  with the standard deviation of $\operatorname{Stdev}\big[ p_U (w)\big] = 0.125 p_U(0)$. We also assume that $\epsilon_P=\epsilon_Q=\epsilon_V=\epsilon$ and $\epsilon_I = 2.5\epsilon$.

All models are implemented in Julia using JuMP \cite{juliajump} and JuMPChance \cite{jumpchance}; our code can be downloaded in \cite{julia_code_gist}. \textcolor{black}{We use the MatpowerCases package~\cite{matpower_cases} to access  the system data.}
The solution of the AC CC-OPF \eqref{approximateACCCOPF} \textcolor{black}{using Gurobi} is compared against the deterministic AC OPF \eqref{detACOPF}, which is solved using Ipopt. As described in Section~\ref{sec:ac_pf_eq_linearization}, the solution of \eqref{detACOPF} is also used as the linearization point for \eqref{approximateACCCOPF}. 

\begin{table}[h!]

\captionsetup{justification=centering, labelsep=period, font=footnotesize, textfont=sc}
\caption{Hourly Wind Power Forecasts (MW) }

\begin{center}
\begin{tabular}{ c p{0.2cm} p{0.2cm} p{0.2cm} p{0.2cm} p{0.2cm} p{0.2cm} p{0.2cm} p{0.2cm} p{0.2cm} p{0.2cm} p{0.2cm}  }
\hline\hline
Node \#& 3 & 8 & 11 & 20 &  24 & 26 & 31& 38 & 43 & 49 & 53  \\
\hline
$p_U(0)$ & 70 & 147  & 102& 105 & 113& 84 & 59 & 250 & 118 & 76 & 72 \\
\hline\hline 
\end{tabular}
\end{center}

\label{table:data}
\end{table}

\subsection{Ex-Ante Cost and Computing Times} \label{sec:ex_ante}
Table~\ref{tab:table_1} compares the deterministic and AC CC-OPF solutions in terms of their ex-ante cost of the deterministic and chance constrained formulations for different values of $\epsilon$. The ex-ante cost is the value of the objective function of the respective formulation, i.e. it reflects a hypothesized dispatch cost before uncertain quantities materialize. 
Note that the deterministic AC OPF formulation \eqref{detACOPF} is extended to include the probabilistic reserve requirement as defined by \eqref{eq:reserveCC1} and \eqref{DETReserve} to account for the need to balance uncertain power injections and thus also includes reserve variables $r$.

The deterministic solution, which only accounts for $\epsilon$ when determining the total amount of reserves, but ignores the effect of reserve allocation on, e.g., line flows and voltage magnitudes,
is insensitive to the value of $\epsilon$, while the ex-ante cost of the AC CC-OPF increases as the value of $\epsilon$ decreases. Lower values of $\epsilon$ indicate a  lower tolerance  level to constraint violations  and thus require  more conservative and costly dispatch  decisions. 
\textcolor{black}{Given the results in Table~\ref{tab:table_2}, we observe that the AC CC-OPF is computed within the same time frame as the deterministic case. Note that these results do not intend to represent the fastest possible AC OPF or AC CC-OPF solves, but rather aim to be illustrative of the relative computational burden, up to orders of magnitude. }

\begin{table}
\centering    
\captionsetup{justification=centering, labelsep=period, font=footnotesize, textfont=sc} 
\caption{Ex-Ante Cost ($\cdot 10^3$, \$) of the Deterministic and AC CC-OPF}
\begin{tabular}{p{1.5cm}| p{0.4cm}| p{0.4cm}| p{0.4cm}| p{0.4cm}| p{0.4cm}| p{0.4cm}| p{0.4cm}| p{0.4cm}}
\hline\hline
\multirow{2}{*}{\textcolor{black}{Model}} & \multicolumn{8}{|c}{$\varepsilon =   \varepsilon_P = \varepsilon_Q = \varepsilon_V = \varepsilon_I/2.5$ in \% }   \\
\cline{2-9}
& 20 & 10  & 5 & 1& 0.5 & 0.1 & 0.05 & 0.01 \\
\hline\hline
AC OPF      & 91.1 & 91.1  & 91.1 & 91.1  & 91.1 & 91.1  & 91.1 & 91.1\\
AC CC-OPF   & 91.2 & 91.6  & 92.1 & 93.1  & 93.6 & 94.02 & 95.6 & 101.1 \\
$\Delta$ (\%) & 0.14 & 0.58  & 1.08 & 2.80  & 2.83 & 3.40  & 4.96 & 10.96\\
\hline\hline
\end{tabular} \\
$\Delta$ denotes the difference between the two formulations.
\label{tab:table_1}

\end{table}

\begin{table}
\centering    \color{black}
\captionsetup{justification=centering, labelsep=period, font=footnotesize, textfont=sc} 
\caption{CPU Times (s) of the Deterministic and AC CC-OPF}
\begin{tabular}{p{1.5cm}| p{0.4cm}| p{0.4cm}| p{0.4cm}| p{0.4cm}| p{0.4cm}| p{0.4cm}| p{0.4cm}| p{0.4cm}}
\hline\hline
\multirow{2}{*}{\textcolor{black}{Model}} & \multicolumn{8}{|c}{$\varepsilon =   \varepsilon_P = \varepsilon_Q = \varepsilon_V = \varepsilon_I/2.5$ in \% }   \\
\cline{2-9}
 & 20 & 10  & 5 & 1& 0.5 & 0.1 & 0.05 & 0.01 \\
\hline\hline
AC OPF      & 1.39 & 1.28  & 1.18 & 1.96 & 2.09 & 2.15 & 3.1 & 2.3 \\
AC CC-OPF   & 1.31 & 1.25  & 1.65 & 1.22 & 1.43 & 1.52 & 2.94 & 3.8 \\
\hline\hline
\end{tabular}

\label{tab:table_2}

\end{table}

\subsection{Ex-Post Comparison} \label{sec:expost}

We now evaluate how the AC CC-OPF and deterministic OPF solutions obtained as in Section~\ref{sec:ex_ante} perform for different uncertainty realizations. We sample 1,000  realizations for each wind farm location as  ${p}_{U}(\omega) = p_U + \omega $, where $\omega \sim N(0, \operatorname{Stdev}\big[ p_U (w)\big])$. 
{ \color{black} For each realization, we re-dispatch the AC CC-OPF and OPF deterministic solutions using the following problem:
\begin{subequations} \label{eq:validation_model}
\begin{align}
& \min \sum_{(ij) \in \mathcal{L}} s^L_{ij} + \sum_{i \in \mathcal{G}} \big( \overline{s}^P_i +\underline{s}^P_i + \overline{s}^Q_i +\underline{s}^Q_i  \big) + \sum_{i \in \mathcal{N}} \big( \overline{s}^V_i +\underline{s}^V_i  \big) \\
& \text{Eq.}~\eqref{ac_powerbal}, \eqref{ac_slack} \\
&p_{G,i}^{min} - \underline{s}^P_i  \leq p_{G,i} \leq p_{G,i}^{max} + \overline{s}^P_i , \forall {i\in\mathcal{N}}  \label{ac_p2}\\
&q_{G,i}^{min} - \underline{s}^Q_i  \leq q_{G,i} \leq q_{G,i}^{max} + \overline{s}^P_i , \forall {i\in\mathcal{N}}  \label{ac_q2}\\
&v_{j}^{min} - \underline{s}^V_i \leq v_{j} \leq v_{j}^{max} + \overline{s}^V_i, \forall {j\in\mathcal{N}}  \label{ac_v2}\\
&(f^{p}_{ij}(\theta, v))^2\!\!+\!\!  (f^{q}_{ij}(v, \theta))^2 \!\!\leq\!\! (s_{ij}^{max} + s^L_{ij})^2, \forall {ij\in\mathcal{L}}, \label{ac_s2}   
\end{align}
\end{subequations}
where $s^L_{ij}$, $\overline{s}^P_i$, $\underline{s}^P_i$, $\overline{s}^v_i$, $\underline{s}^V_i$ are non-negative slack variables introduced to  penalize respective constraint violations. In addition to the constraints stated in \eqref{eq:validation_model}, we fix all variables to their values determined by the response policies in Table~\ref{table:response_pol}, except for those variables denoted as implicit. For the AC CC-OPF case,  we use  the values of $\alpha$ obtained directly from \eqref{approximateACCCOPF}. For comparison with the deterministic AC OPF, we compute $\alpha$ using the following two common practices \cite{6555956}:}
\begin{enumerate}
\item \textit{Uniform policy} assumes that generators respond based on the compulsory participation factor, i.e.  $\alpha_i =1/N_G$, where $N_G$ is the number of generators. 
\item \textit{Reserve-based policy} assumes that each generator responds based on its reserve allotment relative to the capacity reserve requirement, i.e. $\alpha_i = r_i/\sum_{i \in N_G} r_i$.
\end{enumerate}
To facilitate a more direct comparison, we also solve and then re-dispatch the AC CC-OPF with the uniform policy $\alpha_i =1/N_G$, i.e.  $\alpha_i$ is a parameter, not a decision variable.



\textit{1) Feasibility:} We now compare the feasibility of the deterministic and AC CC-OPF solutions  with different response policies. \textcolor{black}{A solution is considered feasible if it can be redispatched for a given uncertainty realization to fully counteract the uncertainty realization without inducing additional constraint violations. If the dispatchtable resources are not sufficient, i.e. slack variables in \eqref{ac_p2} are non-zero, the solution is considered as infeasible. To characterize this infeasibility numerically, imbalance metrics $\overline{I} = \sum_{i \in \mathcal{G}} \overline{s}_i^P$ and $\underline{I} = \sum_{i \in \mathcal{G}} \underline{s}_i^P$ are computed for each realization. Among 1,000 realizations studied, it was systematically observed that $\underline{I}=0$ for the AC CC-OPF solution. The average magnitude of $\overline{I}$ is summarized  in Table~\ref{tab:table_3} and compared with the deterministic AC OPF.} 
The value of this mismatch metric remains constant for the deterministic AC OPF, but monotonically reduces for the AC CC-OPF solution as the value of parameter  $\epsilon$ reduces. Furthermore, using the optimized response policy for redispatching is more advantageous than the uniform response policy. 

\begin{table}[!b] \color{black}
\centering    
\captionsetup{justification=centering, labelsep=period, font=footnotesize, textfont=sc} 
\caption{Average value of metric $\overline{I}$ (MW) characterizing violations of the $p_{G,i}^{max}$ limit in \eqref{CCgen}}
\begin{tabular}{c| c| c| c| c}
\hline\hline
\multirow{2}{*}{\textcolor{black}{$\epsilon^{*}$ in \%}}   &  \multicolumn{2}{c|}{AC OPF }  &  \multicolumn{2}{c}{AC CC-OPF} \\
\cline{2-5}
  & Uniform & Reserve & Uniform & Optimized \\
\hline\hline
20   & 6.1 & 7.6 & 4.8 & 4.6 \\
10   & 6.1 & 7.6 & 4.3  & 4.1\\
5    & 6.1& 7.6 & 2.6& 2.1\\
1    & 6.1 & 7.6 & 0.7& 0.4\\
0.5  & 6.1 & 7.6 & 0.3& 0.2\\
0.1  & 6.1 & 7.6 & 0.1& 0.1\\
0.05 & 6.1 & 7.6 & 0.1& 0.1\\
0.01 & 6.1 & 7.6 & 0.1& 0.1\\
\hline\hline
\end{tabular} \\
Note that $\varepsilon =   \varepsilon_P = \varepsilon_Q = \varepsilon_V = \varepsilon_I/2.5$.
\label{tab:table_3}

\end{table}

\begin{figure}[!t]
  \centering
  \includegraphics[width=\columnwidth]{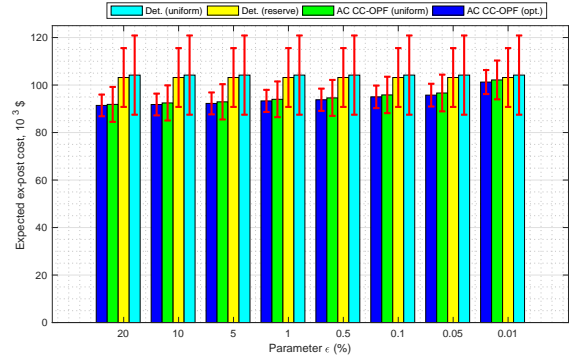}
  \caption{Average ex-post cost and its standard deviation of the deterministic and AC CC-OPF formulations with different response policies for feasible uncertainty realizations.}
  \label{fig:exante}

\end{figure}

\textit{2) Ex-Post Cost:} Figure~\ref{fig:exante} compares the average ex-post cost and corresponding standard deviation for each formulation and varying choices of $\epsilon$. The AC CC-OPF solution has a lower average cost than the deterministic OPF, and also a lower standard deviation, which is an indicator of its  robustness to random uncertainty realizations. The optimized response policy yields lower standard deviations than the uniform and reserve-based response policies.


\textit{3) Analysis of Constraint Violations: } \textcolor{black}{This section analyzes violations on constraints  \eqref{quadapproxsoc} and \eqref{CCv} for the case with the AC CC-OPF with optimized $\alpha$, which would occur if slack variables   $s^L_{ij}$, $\overline{s}^V_{ij}$ and  $\underline{s}^V_{ij}$  attain non-zero value when \eqref{eq:validation_model} is solved for a particular uncertainty realization.}  
\textcolor{black}{In the following results, the frequency of constraint violations at lines \#119 exhibits abnormal behavior for relatively large values of $\epsilon$. This abnormality is dealt with in Section~\ref{sec:line119}.}


Empirical violations of \eqref{quadapproxsoc} are itemized for each transmission line in Table~\ref{tab:table_lineviol2}. However, since the AC CC-OPF model relies on the linear approximation of AC power flows, this probability guarantee is not expected to strictly hold for nonlinear, nonconvex AC power flows used  in \eqref{eq:validation_model}. As expected, tightening chance constraints by reducing the value of parameter $\epsilon_I$ lowers the frequency of violations, or even eliminates the violations entirely for some lines. On the other hand, as the value of parameter $\epsilon_I$ reduces, empirical violations tend to exceed the expected threshold due to the inaccuracies induced by the power flow linearization in \eqref{ac_lin_jacob}. For example, lines \#25 and \#119 have their power flow limits violated in 0.1\% and 0.7\% of cases instead of the postulated value of  $\epsilon_I = 0.01\%$.  In general, these discrepancies are notable for lower values of parameter $\epsilon_I$, which yield a more conservative AC CC-OPF solution that departs farther away from the linearized operating point in \eqref{ac_lin_jacob}. However, the magnitude of probability guarantee  violations due to linearization errors is on par with $\epsilon_I$ and, therefore, is deemed acceptable. 


\begin{table}[t!]
\centering    \color{black}
\captionsetup{justification=centering, labelsep=period, font=footnotesize, textfont=sc} 
\caption{\color{black}Empirical violations of constraint \eqref{quadapproxsoc} (\%)}
\begin{tabular}{c| c| c| c| c| c| c| c| c}
\hline\hline
\multirow{2}{*}{\textcolor{black}{Line ind}} & \multicolumn{8}{c}{$\varepsilon_I =2.5 \varepsilon$  in \% }   \\
\cline{2-9}
 & 20 & 10  & 5 & 1& 0.5 & 0.1 & 0.05 & 0.01 \\
\hline\hline
8    & 0.6    &	1     &	0.1 & 0.3 & 0.2 & 0     & 0     & 0\\
12   & 1.2    &	10.4  &	0.2 & 0.1 & 0.1 & 0     & 0     & 0\\
25   & 18.5   &	11    &	5.4 & 1.2 & 0.6 & 0.2   & 0.2   & 0.1 \\
26   & 0      &	0     &	0   & 0   & 0.2 & 0.2   & 0.1   & 0\\
37   & 15.8   & 9.7	  &	4.5 & 0.6 & 0.3 & 0.2   & 0.2   & 0\\
41   & 19.4   &	15    &	12.1& 4.4 & 2.3 & 0.3   & 0.1   & 0\\
51   & 12.9   &	3.2   &	0.3 & 0   & 0   & 0     & 0     & 0\\
52   & 26     &	16.8  &	9.2 & 3.3 & 1.1 & 0.2   & 0     & 0\\
54   & 11.7   &	4     &	1.9 & 0.4 & 0.2 & 0     & 0     & 0\\
60   & 0      &	0.1   &	0.1 & 0.1 & 0.1 & 0.1   & 0.1   & 0\\
74   & 9.6    &	9     &	4.1 & 0.6 & 0   & 0     & 0     & 0\\
119  & 90.3   &	98.7  &	100 & 100 & 100 & 5.7   & 1.2   & 0.7 \\
\hline\hline
\end{tabular}
\label{tab:table_lineviol2} \\
\end{table}

\textcolor{black}{Among voltage limits in \eqref{CCv}, the $v_i^{min}$ limit is never violated and violations are only observed for the  $v_i^{max}$ limit.  The latter violations are itemized in  Table~ \ref{tab:table_volviol2}. } As the value of parameter $\epsilon$ reduces, so does the frequency of violations observed under the AC CC-OPF solution. Tightening chance constraints  also reduces the number of nodes where violations of \eqref{CCv}  are observed. Since \eqref{CCv} are linear chance constraints, their reformulation is exact and does not use any approximation, unlike in \eqref{quadapproxsplit}-\eqref{absapprox}. As a result, the empirical constraint violations in Table~ \ref{tab:table_volviol2} comply with the probability guarantee imposed by $\epsilon$.

%
%

\begin{table}[t!]
\centering    
\color{black}
\captionsetup{justification=centering, labelsep=period, font=footnotesize, textfont=sc} 
\caption{\color{black} Empirical violations of the $v_i^{max}$ limit in \eqref{CCv}, \%.}
\begin{tabular}{c| c| c| c| c| c| c| c| c}
\hline\hline
\multirow{2}{*}{\textcolor{black}{Node ind}} & \multicolumn{8}{|c}{$\varepsilon_V =\varepsilon$   in \% }   \\
\cline{2-9}
 & 20 & 10  & 5 & 1& 0.5 & 0.1 & 0.05 & 0.01 \\
\hline\hline
3    & 7.4      &	5.1     &	1.5   & 0.1   & 0& 0   & 0   & 0\\
11   & 5.3      &	4.4     &	0.9   & 0.2   & 0.2& 0   & 0   & 0\\
43   & 8.1      &	1.9     &	0.2   & 0   & 0& 0   & 0   & 0\\
53   & 4.2      &	2.8     &	0.2   & 0   & 0& 0   & 0   & 0\\
\hline\hline
\end{tabular}
\label{tab:table_volviol2}
\end{table}

{\color{black}
\subsection{Investigation of violations on line \#119} \label{sec:line119}

\begin{table}[t!]
\centering    \color{black}
\captionsetup{justification=centering, labelsep=period, font=footnotesize, textfont=sc} 
\caption{ \color{black} Empirical violations of constraints \eqref{quadapproxsoc} (\%) with the apparent power flow limit on line 119 set to 90\% of its nominal value.}
\begin{tabular}{c| c| c| c| c| c| c}
\hline\hline
\multirow{2}{*}{\textcolor{black}{Line ind}} & \multicolumn{6}{|c}{$\epsilon_I$ in \% }   \\
\cline{2-7}
 & 20 & 10  & 5 & 1& 0.5 & 0.1  \\
\hline\hline
8   & 0.7       &	3.8  &	0.1 & 0.3 & 0.2 & 0 \\
12  & 1.1       &	10.4 &	5   & 0.1 & 0.1 & 0  \\
25  & 18.6      &	10.7 &	5.5 & 1.2 & 0.6 & 0.2  \\
26  & 0         &   0    &	0   & 0   & 0.2 & 0.2 \\
37  & 15.7      &	9.7  &	4.5 & 0.6 & 0.3 & 0.2  \\
41  & 20.8      &	14.5 &	10.5& 3.5 & 2   & 0.3 \\
51  & 12.6      &	6.4  &  1.5 & 0   & 0   & 0 \\
52  & 25.1      &	12.5 &	6.9 & 1.3 & 1.1 & 0.1 \\
54  &12.2       &   2.9  &	1.5 & 0.4 & 0.2 & 0  \\
60  & 0         &	0.1  &  0.1 & 0.1 & 0.1 & 0.1\\
74  & 17.3      &	9    &	4.1 & 0.6 & 0   & 0  \\
183 & 85.1      &	52.0 &	37.4& 13.3& 10.7& 0.1 \\
\hline\hline
\end{tabular}
\label{table:119_viol} \\
\end{table}

\begin{table}[t!]
\color{black}
\centering    
\captionsetup{justification=centering, labelsep=period, font=footnotesize, textfont=sc} 
\caption{\color{black}Average Ex-Post Cost of the AC CC-OPF with Different Power Flow Limits on Line \#119.}
\begin{tabular}{c| c| c| c| c| c| c}
\hline\hline
\multirow{2}{*}{\textcolor{black}{Limit}} & \multicolumn{6}{|c}{$\epsilon_I$ in \% }   \\
\cline{2-7}
 & 20 & 10  & 5 & 1& 0.5 & 0.1  \\
\hline\hline
Nominal   & 91,421   &	91,803 &	92,260 & 93,313& 93,803& 95,030 \\
90\%   & 91,508  &	91,886 &	92,345 & 94,405& 93896,2 & 95,137 \\
\hline\hline
\end{tabular}
\label{table:119_cost} 
\end{table}

Recall that the empirical frequency of constraint violations on line \#119 reported in Section~\ref{sec:expost} substantially exceeds the value of $\varepsilon_I$ chosen for the AC CC-OPF in \eqref{approximateACCCOPF}. This motivates our investigation of underlying reasons. Line \#119 is directly connected to the reference node, which  accumulates the residual power mismatch across the entire network because it is the only node for which  $p_{G,i}$ is not fixed in \eqref{eq:validation_model}. Given the linearization, the response policy in Table~\ref{table:response_pol} implies a unique value $\bar{p}_{ref}$ for $p_{G,i}$ at the reference node, which differs from the physically correct value  $\hat{p}_{ref}$  computed using the nonlinear, nonconvex deterministic AC OPF model in \eqref{eq:validation_model}. Comparing the two cases,  we observe that $\bar{p}_{ref}$ systematically underestimates $\hat{p}_{ref} $. This mismatch between $\overline{p}_{ref} $ and  $\hat{p}_{ref} $ is attributable only to the linearization. Since line \#119 has a binding apparent power flow limit constraint in the AC CC-OPF, extra power injected at the reference node due to $\overline{p}_{ref} < \hat{p}_{ref} $ leads to  an overload of line \#119. The consistent direction of the observed mismatch is not immediately explained by the models used in this paper and merits further investigation.

We account for the systematic difference in $\overline{p}_{ref} $ and  $\hat{p}_{ref} $ using expert judgment, which is commonly used in power system operations~\cite{Lipka2017}, and manually update given power flow limits. To this end, the power flow limit on line \#119 is reduced to 90\% of its nominal value used in prior experiments to obtain a more conservative solution and, thus, to avoid the violations. As a result, the average ex-post cost of this more conservative solution moderately increases as shown in Table~\ref{table:119_cost}. At this expense, the empirical violations of the apparent power flows reduce as shown in Table~\ref{table:119_viol}. Note that we again observe higher than expected violations on line \#183, which is one line removed from the reference node. However, these violations are smaller than the original violations on line \#119 and are more effectively mitigated by setting the value of parameter $\varepsilon_I$ to a smaller number. 

In practice, the system operator is expected to use the proposed AC CC-OPF with well-defined power flow limits and therefore such abnormalities are unlikely. Otherwise, they can be resolved by refining power flow limits as shown above. }

\section{Conclusion}

This paper describes an AC CC-OPF that accounts for both the AC power flows and the stochasticity of renewable generation resources.  The AC power flows are linearized using the first-order Taylor expansion, which makes it possible to introduce and analytically reformulate chance constraints on generation output, line apparent power flow and nodal voltage limits. The reformulated AC CC-OPF is an SOCP, which is computationally tractable even for large systems. Our case study shows that despite the use of some assumptions and approximations, the AC CC-OPF is able to meet the probabilistic security criteria and is as computationally tractable as its deterministic benchmark. 

\section*{Acknowledgements}
The authors would like to thank Scott Backhaus and Michael Chertkov for the discussions and initial ideas that helped shape this work. The authors are further grateful to the Center for Nonlinear Studies (CNLS) at Los Alamos National Laboratory for providing financial support to Line Roald and to Russell Bent for hosting Miles Lubin as a visitor at CNLS. The work of Yury Dvorkin was in part supported by NSF Award \# CMMI-1825212.

	\bibliographystyle{IEEEtran}
	\bibliography{ref_ccacopf}

\end{document}